\documentclass[english,aps,preprint,nofootinbib]{revtex4}
\usepackage[T1]{fontenc}
\usepackage[latin9]{inputenc}
\usepackage{amsmath}
\usepackage{amssymb}
\usepackage{graphicx}

\makeatletter

\providecommand{\tabularnewline}{\\}

\@ifundefined{textcolor}{}
{%
 \definecolor{BLACK}{gray}{0}
 \definecolor{WHITE}{gray}{1}
 \definecolor{RED}{rgb}{1,0,0}
 \definecolor{GREEN}{rgb}{0,1,0}
 \definecolor{BLUE}{rgb}{0,0,1}
 \definecolor{CYAN}{cmyk}{1,0,0,0}
 \definecolor{MAGENTA}{cmyk}{0,1,0,0}
 \definecolor{YELLOW}{cmyk}{0,0,1,0}
}

\makeatother

\usepackage{babel}
\begin{document}

\preprint{BROWN-HET-1646}

\title{Hidden Kerr/CFT at finite frequencies}

\author{David A. Lowe}

\email{lowe@brown.edu}

\selectlanguage{english}%

\affiliation{Physics Department, Brown University, Providence, RI 02912, USA}

\author{Ilies Messamah}

\email{Ilies.Messamah@wits.ac.za}

\selectlanguage{english}%

\affiliation{National Institute for Theoretical Physics, School of Physics and
Centre for Theoretical Physics, University of the Witwatersrand, Wits
2050, South Africa}

\author{Antun Skanata}

\email{antun_skanata@brown.edu}

\selectlanguage{english}%

\affiliation{Physics Department, Brown University, Providence, RI 02912, USA}
\begin{abstract}
Massless fields propagating in a generic Kerr black hole background
enjoy a hidden $SL(2,\mathbb{R})\times SL(2,\mathbb{R})$ symmetry.
We determine how the exact mode functions decompose into representations
of this symmetry group. This extends earlier results on the low frequency
limit of the massless scalar case to finite frequencies and general
spin. As an application, we numerically determine the parameters of
the representations that appear in quasinormal modes. These results
represent a first step to formulating a more precise mapping to a
holographic dual conformal field theory for generic black holes.
\end{abstract}
\maketitle

\tableofcontents{}

\pagebreak{}

\section{Introduction}

It is a curious result that the entropy formula for a generic Kerr
black hole can be simply explained using a two-dimensional conformal
field theory at finite temperature as a holographic dual \cite{Castro:2010fd}.
This has motivated attempts to determine the properties of such a
conformal field theory. In the near extremal-limit there has been
some success in this direction \cite{Guica:2008mu,Bredberg:2009pv,Cvetic2009,Guica2011a,Detournay2012}.
In addition, results have been claimed in the low frequency limit
of generic Kerr. A review of these various approaches is \cite{Compere:2012jk}.
Clearly this is a very important problem, as a complete description
of the holographic theory could lead to an exact quantum description
of black holes beyond the semiclassical limit commonly studied, or
the exact descriptions found in special limits in string theory.

In the present work we study in more detail the symmetry structure
of the equations of motion for massless fields of general spin in
a generic Kerr background without taking any other special limits.
We find a hidden $SL(2,\mathbb{R})\times SL(2,\mathbb{R})$ symmetry
structure which matches that of the global conformal group of a two-dimensional
conformal field theory. 

In order to construct the parameters that characterize the representations
that appear, it is necessary to solve an eigenvalue problem that may
be expressed as a continued fraction equation. The solutions to this
problem are exhibited as a low frequency expansion, and also computed
numerically for quasinormal modes. The highly damped quasinormal modes
are thought to have a close connection to the exact description of
black hole entropy, and hence the holographic conformal field theory.
We develop the numerical solution of the eigenvalue problem in this
limit to determine the associated representations. 

The Kerr mode functions lead to non-unitary representations of $SL(2,\mathbb{R})\times SL(2,\mathbb{R})$
for the mode functions which reflect the non-invariance of the coordinate
patch under this group. However the representations that appear match
exactly what one expects of the BTZ black hole, for which the correspondence
between gravity and a CFT at finite left/right temperatures is well-understood
\cite{Birmingham:2001pj}. These results provide useful clues and
constraints on the structure of the holographic dual to a generic
Kerr black hole.

\section{$SL(2,\mathbb{R})\times SL(2,\mathbb{R})$ and massless fields in
Kerr\label{sec:-and-Massless}}

The equations of motion of a massless field in the Kerr geometry can
be organized using the symmetry group $SL(2,\mathbb{R})\times SL(2,\mathbb{R})$.
To see this, we use Teukolsky's separation of the equation of motion
\cite{Teukolsky:1972my} in Boyer-Lindquist coordinates:
\[
ds^{2}=\frac{\Delta}{\rho^{2}}\left(dt-a\sin^{2}\theta d\phi\right)^{2}-\frac{\sin^{2}\theta}{\rho^{2}}\left((r^{2}+a^{2})d\phi-adt\right)^{2}-\frac{\rho^{2}}{\Delta}dr^{2}-\rho^{2}d\theta^{2}\,,
\]
defining $\Delta=r^{2}-2Mr+a^{2}$, $\rho^{2}=r^{2}+a^{2}\cos^{2}\theta$.
The solution takes the form, for angular quantum numbers $\ell$ and
$m$
\[
\psi=e^{-i\omega t}e^{im\phi}S_{\ell}^{m}(\theta)R_{\omega\ell m}(r)\,,
\]
with $S_{\ell}^{m}$ being a spin weighted spheroidal harmonic, dependent
on the spin weight $s$. This satisfies the angular equation
\begin{equation}
\left(\frac{d}{dy}(1-y^{2})\frac{d}{dy}+\frac{1}{4}q^{2}\epsilon^{2}y^{2}-sq\epsilon y-\frac{m^{2}+s^{2}+2msy}{1-y^{2}}+E\right)S_{\ell}^{m}(y)=0\,,\label{eq:angular}
\end{equation}
with angular eigenvalue $E$. The radial equation takes the form
\begin{equation}
\left(\Delta^{-s}\frac{d}{dr}\Delta^{s+1}\frac{d}{dr}+\left[V_{0}(r)+V_{\omega\ell m}(r)\right]\right)R_{\omega\ell m}(r)=0\,,\label{eq:radial}
\end{equation}
where the potentials $V_{0}$ and $V_{\omega\ell m}$ are defined
as
\begin{eqnarray*}
V_{0}(r) & = & \frac{r_{+}-r_{-}}{r-r_{+}}\omega_{+}(\omega_{+}-is)-\frac{r_{+}-r_{-}}{r-r_{-}}\omega_{-}(\omega_{-}+is)\\
V_{\omega\ell m}(r) & = & s(s+1)-E+\epsilon(\epsilon-is)+r^{2}\omega^{2}+r\omega(\epsilon+2is)\,,
\end{eqnarray*}
and we have introduced
\[
\omega_{\pm}=\frac{2Mr_{\pm}\omega-am}{r_{+}-r_{-}},\quad\epsilon=2M\omega,\quad q=\frac{a}{M},\quad y=\cos\theta\,.
\]

Around a low frequency limit the radial equation solutions may be
expanded in terms of hypergeometric functions \cite{Mano:1996gn,Mano:1996vt}

\begin{eqnarray}
R_{\omega\ell m}(x) & = & e^{i\epsilon\kappa x}(-x)^{-s-\frac{i}{2}\left(\epsilon+\tau\right)}(1-x)^{\frac{i}{2}\left(\epsilon-\tau\right)}\label{eq:manoexpan}\\
 &  & \sum_{n=-\infty}^{\infty}a_{n}^{\nu}\,\,_{2}F_{1}\left(n+\nu+1-i\tau,-n-\nu-i\tau;1-s-i\epsilon-i\tau;x\right)\,,\nonumber 
\end{eqnarray}
where the solution is chosen to satisfy ingoing boundary conditions
at the horizon. Here we find it convenient to use $x=\omega(r_{+}-r)/\epsilon\kappa$,
$\kappa=\sqrt{1-q^{2}}$, $\tau=(\epsilon-mq)/\kappa$. The parameter
$\nu$ will ultimately determine the $SL(2,\mathbb{R})\times SL(2,\mathbb{R})$
representations that appear in the mode function, and will be determined
momentarily. In a low frequency expansion, $\nu=\ell+\mathcal{O}(\epsilon^{2})$.
For sufficiently small $\epsilon$ usually a single term in the expansion
survives at leading order. Remarkably, it was found the series converges
for all $r<\infty$ beyond the small $\epsilon$ limit.

If we define
\begin{equation}
B_{\omega\ell m}(x)=(-x)^{s/2}(1-x)^{s/2}e^{-i\kappa\epsilon x}R_{\omega\ell m}^{\nu}(x)\,,\label{eq:hyperrep}
\end{equation}
we transform the radial equation into
\begin{eqnarray}
\left(\frac{d}{dx}\Lambda\frac{d}{dx}+\left[\frac{\hat{\omega}_{+}^{2}}{x}+\frac{\hat{\omega}_{-}^{2}}{1-x}-\nu(\nu+1)\right]\right)B_{\omega\ell m} & =\nonumber \\
=-\left[2i\kappa\epsilon\Lambda\frac{d}{dx}-i\kappa\epsilon(s-1)\frac{d\Lambda}{dx}-\epsilon^{2}\kappa\frac{d\Lambda}{dx}-E+\nu(\nu+1)+\frac{\epsilon^{2}}{4}\left(7+\kappa^{2}\right)\right]B_{\omega\ell m}\label{eq:qform}
\end{eqnarray}
where
\[
\Lambda=-x(1-x),\qquad\hat{\omega}_{\pm}=\frac{s}{2}\pm i\omega_{\pm}\,.
\]
Now the first line of \eqref{eq:qform} is (up to a trivial modification)
what is known as the $q$-form of the hypergeometric equation. This
form is useful for exhibiting the $SL(2,\mathbb{R})$ structure of
the solution. There are two ways to rewrite this first line as a quadratic
Casimir of $SL(2,\mathbb{R})$, so all together we find a $SL(2,\mathbb{R})\times SL(2,\mathbb{R})$
structure. The generators are

\begin{eqnarray}
J^{\pm} & = & e^{\pm2\pi T_{R}\phi}\left[-\sqrt{\Lambda}\partial_{x}\pm\frac{M}{T_{R}}\frac{T_{L}\Lambda'-T_{R}}{\sqrt{\Lambda}}\partial_{t}\pm\frac{1}{4\pi T_{R}}\frac{\Lambda'}{\sqrt{\Lambda}}\partial_{\phi}\pm\frac{s}{2\sqrt{\Lambda}}\right]\label{eq:L}\\
J^{3} & = & 2M\frac{T_{L}}{T_{R}}\partial_{t}+\frac{1}{2\pi T_{R}}\partial_{\phi}\nonumber 
\end{eqnarray}
and

\begin{eqnarray}
\bar{J}^{\pm} & = & e^{\pm(2\pi T_{L}\phi-t/2M)}\left[-\sqrt{\Lambda}\partial_{x}\pm\frac{M}{T_{R}}\frac{T_{L}-T_{R}\Lambda'}{\sqrt{\Lambda}}\partial_{t}\pm\frac{1}{4\pi T_{R}}\frac{1}{\sqrt{\Lambda}}\partial_{\phi}\pm\frac{s\Lambda'}{2\sqrt{\Lambda}}\right]\label{eq:Lbar}\\
\bar{J}^{3} & = & -2M\partial_{t}+s\nonumber 
\end{eqnarray}
where we have introduced the parameters 
\begin{equation}
T_{R}=\frac{r_{+}-r_{-}}{4\pi a},\qquad T_{L}=\frac{r_{+}+r_{-}}{4\pi a}\,.\label{eq:temps}
\end{equation}
These generators satisfy the $SL(2,\mathbb{R})$ algebra in the form

\begin{eqnarray*}
\left[J^{+},J^{-}\right] & = & 2J^{3}\\
\left[J^{\pm},J^{3}\right] & = & \mp J^{\pm}\,,
\end{eqnarray*}
and likewise for barred generators. The barred and unbarred generators
commute. This generalizes the proposed algebra of \cite{Castro:2010fd}
to general spin weight $s$, up to algebra isomorphism. The Casimir
operators

\[
C=J^{+}J^{-}+J^{3}J^{3}-J^{3},\qquad\bar{C}=\bar{J}^{+}\bar{J}^{-}+\bar{J}^{3}\bar{J}^{3}-\bar{J}^{3}
\]
each reproduce the term in \eqref{eq:qform}
\[
C=\bar{C}=\frac{d}{dx}\Lambda\frac{d}{dx}+\frac{\hat{\omega}_{+}^{2}}{x}+\frac{\hat{\omega}_{-}^{2}}{1-x}\,.
\]
It is worth pointing out that differential generators \eqref{eq:L}
and \eqref{eq:Lbar} appear in the literature on the relation between
representations of $SL(2,\mathbb{R})$ and hypergeometric functions
\cite{1968lie}. A related $SU(2)$ algebra may also be defined for
the angular equation, and was used in \cite{Fackerell1977} to give
a compact derivation of the Press-Teukolsky identities \cite{Teukolsky:1974yv}.
All together, the expansion for a mode function in the Kerr geometry
exhibits a hidden $SL(2,\mathbb{R})\times SL(2,\mathbb{R})\times SU(2)\times SU(2)$
symmetry, which curiously matches the near-horizon isometry symmetry
of the string theory black holes for which the microscopic entropy
counting is well-understood.

From \eqref{eq:manoexpan} we observe $R_{\omega\ell m}(x)$ is invariant
under the exchange $\nu\rightarrow-\nu-1$, $n\rightarrow-n$. In
order to build representations of $SL(2,\mathbb{R})$ under the action
of \eqref{eq:L} and \eqref{eq:Lbar}, we decompose $B_{\omega\ell m}(x)$
into two independent modes:

\[
B_{\omega\ell m}(x)=\sum_{n=-\infty}^{\infty}\left[\tilde{a}_{n}^{\nu}B_{\omega\ell m}^{n+\nu}(x)+\tilde{a}_{n}^{-\nu-1}B_{\omega\ell m}^{n-\nu-1}(x)\right]\,,
\]
where

\begin{eqnarray}
B_{\omega\ell m}^{n+\nu}(x) & = & (-x)^{n+\nu-\frac{s}{2}-\frac{i}{2}\left(\epsilon-\tau\right)}(1-x)^{\frac{s}{2}+\frac{i}{2}\left(\epsilon-\tau\right)}\label{eq:irreps}\\
 &  & \,_{2}F_{1}\left(-n-\nu-i\tau,-n-\nu+s+i\epsilon;-2n-2\nu;1/x\right)\nonumber 
\end{eqnarray}
and 

\[
\tilde{a}_{n}^{\nu}=a_{n}^{\nu}\frac{\Gamma\left(1-s-i\epsilon-i\tau\right)\Gamma\left(2n+2\nu+1\right)}{\Gamma\left(n+\nu+1-i\tau\right)\Gamma\left(n+\nu+1-s-i\epsilon\right)}\,.
\]
The invariance of $B_{\omega\ell m}(x)$ under $\left\{ \nu\rightarrow-\nu-1,n\rightarrow-n\right\} $
is enforced via the assignment $a_{0}^{\nu}=a_{0}^{-\nu-1}$. 

One can define the irreducible $SL(2,\mathbb{R})$ representations
we will be interested in as realized on a set of basis functions $f_{m_{0}}^{u}$
via 

\begin{eqnarray*}
J^{3}f_{m_{0}}^{u} & = & m_{0}f_{m_{0}}^{u}\,,\\
J^{\pm}f_{m_{0}}^{u} & = & (-u\pm m_{0})f_{m_{0}\pm1}^{u}\,.
\end{eqnarray*}
With the help of the hypergeometric identity

\[
\frac{d}{dz}\left[z^{a}F\left(a,b,c;z\right)\right]=az^{a-1}F\left(a+1,b,c;z\right)\,,
\]
where $z=1/x$, we conclude that $e^{-i\omega t+im\phi}B_{\omega\ell m}^{n+\nu}(x)=f_{-i\tau}^{n+\nu}$
under the action of \eqref{eq:L}. For barred generators \eqref{eq:Lbar},
we swap the first two arguments in the hypergeometric function and
use the same identity to show $e^{-i\omega t+im\phi}B_{\omega\ell m}^{n+\nu}(x)=f_{s+i\epsilon}^{n+\nu}$.

In the notation $D\left(u,m_{0}\right)$ of \cite{1968lie}, with
respect to the product algebra $SL(2,\mathbb{R})_{L}\times SL(2,\mathbb{R})_{R}$
the representation associated with the $n$-th term of the expansion
\eqref{eq:hyperrep} is a direct product 
\begin{equation}
D\left(n+\nu,s+i2M\omega\right)_{L}\,\times\, D\left(n+\nu,-i\frac{2M^{2}\omega-am}{\sqrt{M^{2}-a^{2}}}\right)_{R},\label{eq:kerrrep}
\end{equation}
which is a non-unitary representation of the algebra, with Casimir
$(\left|n\right|+\nu)(\left|n\right|+\nu+1)$. Later we will see that
$\nu$ will be real provided $\omega$ is real, but will be complex
for quasinormal modes. With respect to the generators defined above,
the weights of representations are $ik_{L}-2M\omega,\, ik_{R}-\frac{2M^{2}\omega-am}{\sqrt{M^{2}-a^{2}}},$
where $k_{L,R}$ is an integer. The condition that the representations
collapse to a highest weight or lowest weight representation is that
$\nu+i2M\omega$ or $\nu+i\frac{2M^{2}\omega-am}{\sqrt{M^{2}-a^{2}}}$
is an integer. This is generally not satisfied for real non-vanishing
frequencies or momenta. 

The original motivation for the expansion \eqref{eq:manoexpan} was
as a low frequency expansion. As we have seen, this is equivalent
to organizing the expansion according to the $SL(2,\mathbb{R})\times SL(2,\mathbb{R})$
symmetry, which moreover leads to a convergent expansion for general
frequencies%
\footnote{In \cite{Lowe:2011aa} a more general one-parameter family of related
$SL(2,\mathbb{R})\times SL(2,\mathbb{R})$ symmetries was found in
the low frequency limit. In the present work, we are able to treat
the finite frequency regime, which seems to single out the particular
$SL(2,\mathbb{R})\times SL(2,\mathbb{R})$ considered here.%
}. 

This expansion straightforwardly produces low frequency scattering
amplitudes. In particular, the results of Page/Starobinsky \cite{starobinski,Page:1976df}
can easily be recovered by retaining only the $n=0$ term in \eqref{eq:manoexpan},
as shown in \cite{Mano:1996gn}. We discuss this point further in
section \ref{sub:Absorption-probability}.

We will find the parameter $\nu$ becomes a function of frequency,
determined by picking out a convergent solution to \eqref{eq:manoexpan}.
We solve for this parameter in various different limits. But first
we will investigate in more detail the connection to conformal field
theory.

\section{CFT/gravity mapping}

In the above we have shown how a general mode function decomposes
into irreducible representations of the $SL(2,\mathbb{R})\times SL(2,\mathbb{R})$
algebra. To make the meaning of these representations more clear it
is helpful to compare to the analogous computation for the three-dimensional
black hole in asymptotically anti-de Sitter spacetime \cite{Banados:1992gq},
where the same symmetry structure appears, and the holographic dictionary
is well-known.

\subsection{BTZ example}

Quasinormal modes have been studied in this context in \cite{Birmingham:2001pj}
so we find it helpful to follow their notation. The BTZ metric can
be written in the form
\begin{equation}
ds^{2}=\frac{dz^{2}}{4z(1-z)^{2}}+\frac{1}{1-z}\left(-r_{-}dt+r_{+}d\phi\right)^{2}-\frac{z}{1-z}\left(r_{+}dt-r_{-}d\phi\right)^{2}\,,\label{eq:btzmetric}
\end{equation}
where infinity is $z=1$ and the outer horizon sits at $z=0$. Here
$r_{+}$ and $r_{-}$ are the radii of the outer and inner horizons.
These are related to left and right temperatures in the CFT via \eqref{eq:temps}
by replacing $a\to1/2$:

\begin{equation}
T_{R}=\frac{r_{+}-r_{-}}{2\pi},\qquad T_{L}=\frac{r_{+}+r_{-}}{2\pi}\,.\label{eq:BTZtemps}
\end{equation}

For a scalar field of mass $\tilde{m}$, or a vector field of mass
$\tilde{m}$ (with spin parameter $s=\pm1$), an analog of the Teukolsky
solution is
\[
\Phi=e^{-ik_{+}x^{+}-ik_{-}x^{-}}B(z)
\]
where
\[
x^{+}=r_{+}t-r_{-}\phi,\qquad x^{-}=r_{+}\phi-r_{-}t
\]
and
\begin{eqnarray*}
k_{+}+k_{-} & = & \frac{\omega-k}{2\pi T_{R}}\\
k_{+}-k_{-} & = & \frac{\omega+k}{2\pi T_{L}}\,.
\end{eqnarray*}
The radial equation takes the hypergeometric form
\begin{equation}
z(1-z)\frac{d^{2}B}{dz^{2}}+(1-z)\frac{dB}{dz}+\left[\frac{k_{+}^{2}}{4z}-\frac{k_{-}^{2}}{4}-\frac{\tilde{m}^{2}+2s\tilde{m}}{4(1-z)}\right]B=0\label{eq:btzhyper}
\end{equation}
with solution
\begin{equation}
B(z)=z^{\alpha}(1-z)^{\beta}\,_{2}F_{1}(a,b;c;z)\label{eq:btzsol}
\end{equation}
with 
\[
\alpha=-\frac{ik_{+}}{2},\qquad\beta=\frac{1}{2}\left(1-\sqrt{1+\tilde{m}^{2}+2\tilde{m}s}\right),
\]

\[
a=\frac{k_{+}-k_{-}}{2i}+\beta,\qquad b=\frac{k_{+}+k_{-}}{2i}+\beta,\qquad c=1+2\alpha\,.
\]
The radial equation \eqref{eq:btzhyper} is of the same form as the
first line in \eqref{eq:qform} with the replacement $B(z)\to(1-z)^{1/2}\tilde{B}(z)$.
However one subtlety we should address is that a given hypergeometric
equation has 24 different equivalent ways of writing the solution
in terms of hypergeometric functions. These different solutions are
related by Kummer transformations. From the viewpoint of the black
hole, these presumably correspond to mode functions in different coordinate
patches. To compare with the form of the solution \eqref{eq:irreps},
where $x$ runs over the range $(-\infty,0)$ we perform a $z\to y=1-1/z$
Kummer transformation which maps the solution \eqref{eq:btzsol} to
\[
B(x)=(-y)^{c-a-b+\beta}(1-y)^{b-\alpha-\beta}\,_{2}F_{1}(c-a,1-a;c+1-a-b;y)\,.
\]

From here we may read off the parameters of the $SL(2,\mathbb{R})\times SL(2,\mathbb{R})$
representation. In the notation $D(\nu,m_{0})$ of \cite{1968lie}
we find
\begin{equation}
D_{L}\left(\beta-1,i\frac{k_{+}-k_{-}}{2}\right)\,\times\, D_{R}\left(\beta-1,-i\frac{k_{+}+k_{-}}{2}\right)\label{eq:btzrep}
\end{equation}
for which the quadratic Casimir is
\[
C=\bar{C}=\frac{\tilde{m}^{2}+2\tilde{m}s}{4}\,.
\]
These are non-unitary representations of $SL(2,\mathbb{R})$'s although
the Casimir agrees with what we expect for a discrete highest weight
irreducible representation of $SL(2,R)$ with conformal weight $h=\beta$.
The non-unitarity may be attributed to the fact that the external
coordinate patch of the BTZ black hole \eqref{eq:btzmetric} is not
invariant under global AdS isometries. When one takes a pure AdS limit,
the coordinates \eqref{eq:btzmetric} only cover one of an infinite
number of patches needed to cover AdS (or more precisely the covering
space of AdS). Thus, while modes on the covering space of AdS transform
as a unitary highest weight representation%
\footnote{This is treated explicitly for $AdS_{4}$ in \cite{Dusedau:1985ue},
and the result for $AdS_{3}$ follows by decomposing the highest weight
representations of $SO(3,2)$ into $SO(2,2)$.%
} of $SL(2,\mathbb{R})\times SL(2,\mathbb{R})$ they are related to
a nontrivial composition of non-unitary representations of modes on
different patches. At the level of the mode functions, the mapping
between the global mode functions and the BTZ patch mode functions
will (and consequently the group representations will) follow analogous
results in \cite{Unruh:1976db} for Rindler versus Minkowski spacetime.
We conclude that the non-unitary representations \eqref{eq:btzrep}
for a single BTZ patch can be viewed as descending from a unitary
highest weight representation of $SL(2,\mathbb{R})\times SL(2,\mathbb{R})$
corresponding to a primary operator in the CFT with conformal dimension
$\beta$.

\subsection{Conjecture for Kerr/CFT}

We conclude that the representations of $SL(2,\mathbb{R})\times SL(2,\mathbb{R})$
for BTZ modes \eqref{eq:btzrep} match exactly those of Kerr \eqref{eq:kerrrep}.
This leads us to conjecture that an exact mode of Kerr may be reconstructed
from some more fundamental primary conformal field of weight $\nu$
and its descendants.

Let us note that the expansion for a Kerr mode \eqref{eq:manoexpan}
may be expressed as a sum from $n=0$ to $\infty$ simply by using
the symmetry of the hypergeometric function under its first two arguments.
This swaps $\nu\to-\nu-1$ and $n\to-n$. As is well-known in the
AdS/CFT correspondence, both terms may be viewed as originating from
correlators of a conformal primary of weight $\nu$ in the presence
of a source term \cite{Witten:1998qj}.

A new issue that arises in Kerr/CFT is that of the proper normalization
of the modes with physical boundary conditions at spatial infinity.
Certainly one may use the expansion \eqref{eq:manoexpan} to compute
scattering correlators on some surface of large fixed $r$, but to
properly impose incoming/outgoing boundary conditions at spatial infinity,
another expansion must be used that is convergent at $r=\infty.$
Such an expansion is crucial for obtaining quasinormal mode boundary
conditions, or computing genuine scattering amplitude computations
from past infinity to future infinity. To achieve this at the level
of mode functions, one must instead expand in terms of a different
set of basis functions. In \cite{Mano:1996gn,Mano:1996vt,Leaver1986}
this is chosen to be a set of Coulomb functions, though other choices
are possible \cite{Leaver:1985ax}. 

From the CFT point of view, one may view the evolution from $r=\infty$
to finite $r$ as a renormalization group flow. However we know little
about the holographic dual of asymptotically flat spacetime from which
the theory is flowing to the Kerr/CFT in the infrared. Without such
a description we cannot formulate the proper boundary conditions in
the holographic dual that would allow us, for example, to reproduce
quasinormal mode frequencies. Some related works that study the problem
of holographic duals to asymptotically flat spacetime include \cite{Dappiaggi:2004kv,Dappiaggi:2005ci,Bagchi:2012xr}

\subsection{Absorption probability\label{sub:Absorption-probability}}

By matching the solutions obtained via an expansion in Coulomb functions
convergent at $r=\infty$ to the expansion \eqref{eq:manoexpan} valid
at all finite $r$, and with the help of Teukolsky-Starobinsky identities
and a few other simplifying relations, \cite{Mano:1996gn} write down
the exact Kerr absorption rate for all finite frequencies:

\begin{eqnarray}
\sigma_{\text{abs}} & = & \left(2\epsilon\kappa\right)^{2\nu+1}\frac{e^{\pi\epsilon}}{\pi}\sinh\pi\left(\epsilon+\tau\right)\frac{D_{\nu}}{\left|N_{\nu}\right|^{2}}\label{eq:absorption}
\end{eqnarray}
where

\begin{eqnarray*}
N_{\nu} & = & 1+\frac{i}{\pi}\left(2\epsilon\kappa\right)^{2\nu+1}\left(-1\right)^{2s}e^{i\pi\nu}\sin\pi\left(\nu-i\tau\right)\left(\frac{\sin\pi\left(\nu-s-i\epsilon\right)}{\sin2\pi\nu}\right)^{2}D_{\nu}\\
D_{\nu} & = & \left|\frac{\Gamma\left(\nu+1-i\tau\right)\Gamma\left(\nu+1-s+i\epsilon\right)\Gamma\left(\nu+1+s+i\epsilon\right)}{\Gamma\left(2\nu+1\right)\Gamma\left(2\nu+2\right)}\right|^{2}d_{\nu}\\
d_{\nu} & = & \left|\sum_{n\leq0}\frac{\left(-1\right)^{n}}{\left(-n\right)!\left(2\nu+2\right)_{n}}\frac{\left(\nu+1+s-i\epsilon\right)_{n}}{\left(\nu+1-s+i\epsilon\right)_{n}}a_{n}^{\nu}\right|^{2}\\
 &  & \times\left|\sum_{n\geq0}\frac{\left(2\nu+1\right)_{n}}{n!}\frac{\left(\nu+1+s-i\epsilon\right)_{n}}{\left(\nu+1-s+i\epsilon\right)_{n}}a_{n}^{\nu}\right|^{-2}\,.
\end{eqnarray*}
In the small frequency approximation, $\epsilon\ll1$, the absorption
rate properly reproduces the Page formula \cite{starobinski,Page:1976df}.
By keeping $\mathcal{O}(\epsilon)$ terms in \eqref{eq:absorption},
we can set $E=\ell(\ell+1)+\mathcal{O}(\epsilon^{2})$ in \eqref{eq:qform}.
Choosing the integer shift in $\nu$ so that $\nu=\ell+\mathcal{O}(\epsilon)$
makes the $n=0$ term the leading term in the series expansion.

To leading order in $\epsilon$, the formula \eqref{eq:absorption}
reduces to:

\[
\sigma_{\text{abs}}^{\epsilon}=\left(2\epsilon\kappa\right)^{2\ell+1}\frac{e^{\pi\epsilon}}{\pi}\frac{\omega_{R}}{2T_{R}}\prod_{k=1}^{\ell}\left[k^{2}+\left(\frac{\omega_{R}}{2\pi T_{R}}\right)^{2}\right]\times\left(1+\mathcal{O}(\epsilon)\right)
\]
where we introduced the frequencies

\[
\omega_{L}=\frac{2M^{2}\omega}{a},\,\,\,\,\,\,\omega_{R}=\frac{2M^{2}\omega}{a}-m\,.
\]
In the $\epsilon\rightarrow0$ limit the dependence on the left-moving
frequency $\omega_{L}\propto\epsilon$ is rather trivial, however
there is a highly non-trivial dependence on the right-moving frequency
which remains finite $\omega_{R}=-m$. This is a familiar behavior
of the effective string absorption cross section for massless bosons
in $N=4$ supergravity \cite{Gubser1997a}. With the representations
surviving this limit we associate conformal weight $h_{R}=\ell+1$,
compatible with results of \cite{Lowe:2011wu} and conformal weight
identification of \cite{Gubser1997a}.

\subsubsection*{Connection with extremal Kerr/CFT}

Encouraged with our findings, here we speculate on a possible connection
with results obtained in the extremal limit $a\rightarrow M$. More
precisely, we search for regime in which we find representations corresponding
to near horizon extreme Kerr (NHEK) modes near the superradiant bound.
In \cite{Bredberg:2009pv} it was found that the absorption probability
of the modes saturating the superradiant bound in the near-extremal
Kerr background corresponds to a thermal CFT 2-point function:

\begin{equation}
\sigma_{\text{ }}\propto T_{H}^{2\beta}e^{\pi m}\sinh\pi\left(m+\frac{\omega-m/2M}{2\pi T_{H}}\right)\left|\Gamma\left(\frac{1}{2}+\beta+im\right)\right|^{4}\left|\Gamma\left(\frac{1}{2}+\beta+i\frac{\omega-m/2M}{2\pi T_{H}}\right)\right|^{2}\label{eq:superradiant}
\end{equation}
where $\beta^{2}=\frac{1}{4}-2m^{2}+\bar{A}_{lm}$, and $\bar{A}_{lm}$
is the angular eigenvalue evaluated for $a\omega=m/2$. 

We do not observe this truncation of \eqref{eq:absorption} for quasinormal
mode excitations (QNM). In terms of the transmission and reflection
coefficients, the QNMs correspond to frequencies at which both $T$
and $R$ develop poles, in such way that $\left|T\right|\approx\left|R\right|$.
There is also another set of modes one can consider, compatible with
the purely outgoing boundary condition at infinity, called total reflection
modes (TRM). These correspond to frequencies at which transmission
coefficient vanishes, making them standing waves at $r=\infty$. We
observe that total reflection modes with exact frequencies given by
\cite{Keshet2008}

\[
\omega_{TRM}=m\Omega-2\pi iT_{H}(n-s),\,\,\,\,\,\,\,\, n\in\mathbb{\mathbb{\mathbb{N}}}
\]
correctly reproduce the near-superradiant frequencies in the extremal
limit. Here $\Omega=\frac{a}{r_{+}^{2}+a^{2}}$ is the angular velocity
at the outer horizon and $T_{H}=\frac{r_{+}-r_{-}}{4\pi\left(r_{+}^{2}+a^{2}\right)}$
is the Hawking temperature. 

Taking the extremal limit on the parameters of our $SL(2,\mathbb{R})$
representations, 

\[
\lim_{a\rightarrow M}i\frac{2M^{2}\omega_{TRM}-ma}{\sqrt{M^{2}-a^{2}}}=n-s
\]
we arrive at an interesting analogy: just as the BTZ scattering amplitudes
for quasinormal mode frequencies reproduce the pole structure of a
CFT 2-point function \cite{Birmingham:2001pj}, so do the hidden Kerr/CFT
representations in the case of extremal total reflection modes. As
$\kappa\rightarrow0$, the denominator in \eqref{eq:absorption} is
exactly equal to 1, and the Kerr absorption cross section reproduces
\eqref{eq:superradiant}, provided we identify $\nu+1$ with $\beta+1/2$. 

The absorption cross section accounts for poles of a chiral 2-dimensional
CFT 2-point function; we suspect the present understanding of NHEK
asymptotic boundary conditions, providing an enhancement to one Virasoro
only, can then be recast in terms of the monodromy analysis in the
highly damped regime.

The fact that the total reflection modes have no bulk degrees of freedom
corroborates with the findings of \cite{Amsel2009}, which makes the
extremal Kerr/CFT a topological theory. As we know how to count the
microscopic degrees of freedom for $AdS_{3}$ quotients at any value
of the angular momentum \cite{Brown1986b}, it is reasonable to assume
same can be achieved in Kerr/CFT. We expect that the two hidden $SL(2,\mathbb{R})$'s
enhance to a full $vir_{L}\times vir_{R}$, with central charges given
by $c_{L}=c_{R}=12J$. As originally hinted in \cite{Castro:2010fd},
the Cardy formula with this value of central charges, together with
temperatures \eqref{eq:temps}, exactly reproduces the classical Bekenstein-Hawking
entropy of the Kerr black hole.

\section{Eigenvalue equation}

The expansion of the mode functions \eqref{eq:manoexpan} converges
only for special values of the parameter $\nu$. To find this parameter
a continued fraction is set up. Similar methods are used to construct
the angular eigenvalue, and the frequency of quasinormal modes. To
proceed, one expresses the radial equation \eqref{eq:qform} as a
three-term recurrence relation \cite{Mano:1996vt}
\begin{equation}
\alpha_{n}a_{n+1}+\beta_{n}a_{n}+\gamma_{n}a_{n-1}=0\label{eq:threeterm}
\end{equation}
where
\begin{eqnarray*}
\alpha_{n} & = & \frac{i\epsilon\kappa(n+\nu+1+s+i\epsilon)(n+\nu+1+s-i\epsilon)(n+\nu+1+i\tau)}{(n+\nu+1)(2n+2\nu+3)}\\
\beta_{n} & = & -\lambda-s(s+1)+(n+\nu)(n+\nu+1)+\epsilon^{2}+\epsilon(\epsilon-mq)\\
 &  & +\frac{\epsilon(\epsilon-mq)(s^{2}+\epsilon^{2})}{(n+\nu)(n+\nu+1)}\\
\gamma_{n} & = & -\frac{i\epsilon\kappa(n+\nu-s+i\epsilon)(n+\nu-s-i\epsilon)(n+\nu-i\tau)}{(n+\nu)(2n+2\nu-1)}
\end{eqnarray*}
and we have defined
\[
\lambda=E-s(s+1)-2ma\omega+a^{2}\omega^{2}\,.
\]
The eigenvalue equation is expressed using the continued fractions
\begin{equation}
R_{n}=\frac{a_{n}}{a_{n-1}}=-\frac{\gamma_{n}}{\beta_{n}+\alpha_{n}R_{n+1}},\qquad L_{n}=\frac{a_{n}}{a_{n+1}}=-\frac{\alpha_{n}}{\beta_{n}+\gamma_{n}L_{n-1}}\,.\label{eq:continuedfrac}
\end{equation}
For general values of $\nu$ the solution to the recurrence relation
\eqref{eq:threeterm} will diverge as $|n|\to\infty$. To avoid this
and find the so-called minimal solution, one must demand that $\nu$
solve an additional eigenvalue equation
\begin{equation}
R_{n}L_{n-1}=1\,.\label{eq:nueqn}
\end{equation}
There is an equivalence of solutions under $\nu\to\nu+k$ where $k$
is an integer (apparently not noticed in \cite{Mano:1996vt,Mano:1996gn}).
One convention, useful for real frequencies, is to choose the integer
shift in $\nu$ such that $E-\nu(\nu+1)$ term on the right-hand side
of \eqref{eq:qform} is minimized, in order that the $n=0$ term tends
to be the leading order term in the expansion. Another convention,
that will be useful when discussing quasinormal modes, will be to
simply shift the real part of $\nu$ into the range $[0,1/2)$ using
the combined symmetries of the expansion under $\nu\to-\nu-1$ and
$\nu\to\nu+k$.

When $\nu$ satisfies \eqref{eq:nueqn}, one finds 
\[
\lim_{n\to\infty}nR_{n}=-\lim_{n\to-\infty}nL_{n}=\frac{i\epsilon\kappa}{2}\,.
\]
This leads to convergence of the series \eqref{eq:manoexpan} for
all finite radii $r$ \cite{Mano:1996vt}.

\subsection{Low frequency expansion}

In a low frequency expansion, the solution of \eqref{eq:nueqn} is
\begin{eqnarray*}
\nu & = & \ell-\epsilon^{2}\frac{\ell(\ell+1)(-11+15\ell(1+\ell))+6\left(-1+\ell+\ell^{2}\right)s^{2}+3s^{4}}{2\ell(1+2\ell)(1+\ell)(-1+2\ell)(3+2\ell)}\\
 &  & +\epsilon^{3}\frac{mq}{\ell(1+\ell)(-1+2\ell)(1+2\ell)(3+2\ell)}\left[5\ell(1+\ell)-3+\frac{s^{2}\left(3\ell(\ell+1)\left(\ell^{2}+\ell-3\right)+11\right)}{(-1+\ell)\ell(1+\ell)(2+\ell)}\right.\\
 &  & \left.+\frac{s^{4}\left(-16+3\ell(1+\ell)+5s^{2}\right)}{(-1+\ell)\ell(1+\ell)(2+\ell)}\right]+\cdots
\end{eqnarray*}
which is obtained by substituting the low frequency expansion for
$E$ of \cite{Fackerell1977} into \eqref{eq:nueqn}. Note for real
frequencies $\nu$ is also real. As we will see in the following,
$\nu$ becomes complex for quasinormal modes, as does the angular
eigenvalue. The first two terms match the expression in \cite{Mano:1996vt}
and the next term is new. As we have seen in section \ref{sec:-and-Massless},
with $\nu$ and $\omega$ given, the irreducible representations that
appear in the mode function are fully determined.

\subsection{Numerical solutions}

The solution of closely related continued fraction eigenvalue equations
has been considered for spheroidal harmonics in \cite{Fackerell1977,flammer}.
Quasinormal modes have been studied using similar techniques in \cite{Leaver:1985ax,Leaver1986}.
In this case one imposes quasinormal mode boundary conditions on solutions
of the radial equation. Imposing the purely outgoing boundary condition
at infinity requires a different expansion of the radial solution
near infinity. This is then matched to the boundary condition that
the mode be purely infalling on the future horizon, leading to a radial
eigenvalue problem that determines the quasinormal mode frequencies. 

One method that is helpful in improving the convergence of such algorithms
has been developed in \cite{Nollert:1993zz}. When one numerically
computes a continued fraction, such as \eqref{eq:continuedfrac},
some cutoff on $n$ is needed. By choosing an initial value for $R_{n_{\text{cutoff}}}$
judiciously, it is possible to improve the convergence of the continued
fraction, as well as its domain of convergence. In our numerical code,
we apply this method at high order in both the determination of the
quasinormal mode frequencies, as well as the determination of the
$\nu$ eigenvalues. The numerical determination of quasinormal modes
has been studied more recently in \cite{Berti:2003jh,Berti:2004um,Berti:2005gp,Berti:2005ys,Berti:2009kk}
and our results for the frequencies match well with those found there.

\subsubsection{Low order quasinormal modes}

In the case of low order quasinormal modes the spheroidal eigenvalue
equation and the frequency equation must be solved simultaneously.
Towards that end we adopt the technique of Leaver \cite{Leaver:1985ax},
with the Nollert improvement of continued fractions \cite{Nollert:1993zz},
for both the radial and spheroidal eigenvalue equation. This is implemented
using a high precision iterative method in Mathematica, as described
in the Appendix.

Once the quasinormal mode frequency is known to high precision, we
determine $\nu$ by using similar numerical methods to solve the equation
\eqref{eq:nueqn} for $n=1.$ In this case both rising and lowering
infinite continued fractions $R_{n}$ and $L_{n}$ need to be truncated
to some $n_{\text{cutoff}}$, where we use Nollert approximation described
in Appendix to estimate the remainder of the continued fraction. 

Some representative results are shown in Tables I and II, for $s=-2$,
$\ell=2$ and $m=0,1$ quasinormal modes. The quasinormal mode frequencies
precisely agree with results presented in \cite{Leaver:1985ax}. Furthermore,
some sample computations of $\nu$ appear in \cite{Zhang2013}, and
we have checked our numerics correctly reproduces those results. We
conventionally normalize $2M=1$. For clarity of presentation we utilize
the invariance under $\nu\rightarrow-\nu-1$ and $\nu\rightarrow\nu+k$,
$k\in\mathbb{\mathbb{Z}}$ to map our $\nu$ numeric values into the
range $0<Re(\nu)<1/2$. 

\begin{table}
\noindent \begin{centering}
\begin{tabular}{c|c|c|c}
$a/M$ & $\omega$ & $A_{lm}$ & $\nu$\tabularnewline
\hline 
$0.0$ & $0.747343-0.177925i$  & $4.000000+0.000000i$  & $0.271625-0.233965i$ \tabularnewline
$0.1$ & $0.748064-0.177796i$  & $3.999309+0.000348i$  & $0.272325-0.234104i$\tabularnewline
$0.2$ & $0.750248-0.177401i$ & $3.997216+0.001395i$ & $0.274456-0.234519i$\tabularnewline
$0.3$ & $0.753970-0.176707i$ & $3.993667+0.003142i$ & $0.278128-0.235204i$\tabularnewline
$0.4$ & $0.759363-0.175653i$ & $3.988560+0.005596i$ & $0.283541-0.236149i$\tabularnewline
$0.5$ & $0.766637-0.174138i$ & $3.981738+0.008757i$ & $0.291024-0.237327i$\tabularnewline
$0.6$ & $0.776108-0.171989i$ & $3.972969+0.012620i$ & $0.301108-0.238673i$\tabularnewline
$0.7$ & $0.788259-0.168905i$ & $3.961901+0.017153i$ & $0.314673-0.240044i$\tabularnewline
$0.8$ & $0.803835-0.164313i$ & $3.947997+0.022256i$ & $0.333269-0.241085i$\tabularnewline
$0.9$ & $0.824009-0.156965i$ & $3.930384+0.027633i$ & $0.359906-0.240780i$\tabularnewline
\end{tabular}
\par\end{centering}

\noindent \begin{centering}
\caption{Numerical results for $s=-2,$ $\ell=2$ and $m=0$ quasinormal modes.}

\par\end{centering}

\end{table}

\begin{table}
\noindent \begin{centering}
\begin{tabular}{c|c|c|c}
$a/M$ & $\omega$ & $A_{lm}$ & $\nu$\tabularnewline
\hline 
$0.0$ & $0.747343-0.177925i$  & $4.000000+0.000000i$  & $0.271625-0.233965i$ \tabularnewline
$0.1$ & $0.760865-0.177597i$ & $3.948480+0.012231i$ & $0.281465-0.236237i$\tabularnewline
$0.2$ & $0.776496-0.176977i$ & $3.893150+0.025197i$ & $0.293393-0.238774i$\tabularnewline
$0.3$ & $0.794661-0.176000i$ & $3.833210+0.038881i$ & $0.307963-0.241559i$\tabularnewline
$0.4$ & $0.815958-0.174514i$ & $3.767570+0.053241i$ & $0.326020-0.244553i$\tabularnewline
$0.5$ & $0.841265-0.172346i$ & $3.694740+0.068178i$ & $0.348933-0.247684i$\tabularnewline
$0.6$ & $0.871937-0.169128i$ & $3.612470+0.083470i$ & $0.379077-0.250847i$\tabularnewline
$0.7$ & $0.910243-0.164170i$ & $3.517150+0.098594i$ & $0.420976-0.254022i$\tabularnewline
$0.8$ & $0.960461-0.155910i$ & $3.402280+0.112173i$ & $0.484170-0.258022i$\tabularnewline
$0.9$ & $1.032583-0.139609i$ & $3.253450+0.119510i$ & $0.409335+0.266455i$\tabularnewline
\end{tabular}
\par\end{centering}

\noindent \centering{}\caption{Numerical results for $s=-2,$ $\ell=2$ and $m=1$ quasinormal modes.}
\end{table}

\subsubsection{Highly damped quasinormal modes}

In the highly damped regime the simultaneous solution of the spheroidal
eigenvalue equation and the frequency equation becomes numerically
unstable. To make progress, we follow the method of \cite{Berti:2004um}
and use a conjectured asymptotic expansion for the high order eigenvalues
of spheroidal equation
\[
E=(2L+1)iq\epsilon/2+\mathcal{O}(\epsilon^{0})
\]
where $L=\min\left(\ell-\left|m\right|,\ell-\left|s\right|\right)$.
We obtain the quasinormal mode frequencies by numerically searching
for a solution of an inverted continued fraction equation, as proposed
by Leaver \cite{Leaver:1985ax}. At high damping, the numeric solutions
can be shown to follow the analytic result $\omega\simeq\omega_{0}+4\pi iT_{0}\left(N+1/2\right)$,
where $\omega_{0},\, T_{0}$ can also be computed within the WKB approximation
\cite{Keshet2008}, and the overtone number $N$ takes integer values. 

We numerically solve for $\nu$ by finding solutions of the equation
quadratic in the continued fractions, \eqref{eq:nueqn}, where we
find better convergence if we shift index $n$ by the imaginary part
of the quasinormal mode frequency. As a rule of thumb, we use the
number of terms in the continued fraction of the order of the overtone
number and implement the Nollert improvement for the remainder of
the fraction. With each overtone we increase computational precision
until the result converges. The numerics appear most stable when a
starting value $\nu\sim\mathcal{O}(N/2)$ is used. As the solutions
to \eqref{eq:nueqn} are determined up to integer shift, $\nu+k$,
and reflection $\nu\rightarrow-\nu-1$, we can always find a solution
such that $Re(\nu)<1/2$. We use this symmetry when we plot our data. 

In figure 1 we show quasinormal mode frequencies at high damping for
different values of $s$, and $m$ with $\ell=2$. For $m$ fixed,
we observe the real part of the frequency converges to the same value
at high overtones, irrespective of spin weight. The results are compatible
with known numerical computations of Berti et al. \cite{Berti:2004um}.

In figure 2 we display our numerical solutions for $s=-2,\,\ell=2$,
and $m=2$ modes. We plot quasinormal mode data as a function of $a/M$
and for fixed overtone number. We find the frequencies approach the
expected asymptotic behavior with increasing overtones. Our data for
the 400th overtone is comparable to the extrapolated asymptotic curve
computed in \cite{Berti:2004um}, with the plot in figure 2 suggesting
the overtones above 240 already give a qualitatively good estimate
of the asymptotic regime. The improvement in convergence for $\nu$
equation happens roughly around this value of frequency. The plot
of $\nu$ values displays two strands of solutions corresponding to
even and odd valued overtones, which persist at all values of quasinormal
frequency. This arises from the approximate half-integer spacing between
frequencies at high damping, analytically computed in \cite{Keshet2008},
and also numerically confirmed in \cite{Berti:2004um}. In figure
3 we compare $\nu$ values corresponding to a single strand, for fixed
$\ell=2$ and independently varied $s=0,\,-2$ and $m=1,\,2$, where
we map out the solutions of \eqref{eq:nueqn} up to 1000th overtone. 

\begin{figure}
\noindent \begin{centering}
\includegraphics[width=0.45\textwidth]{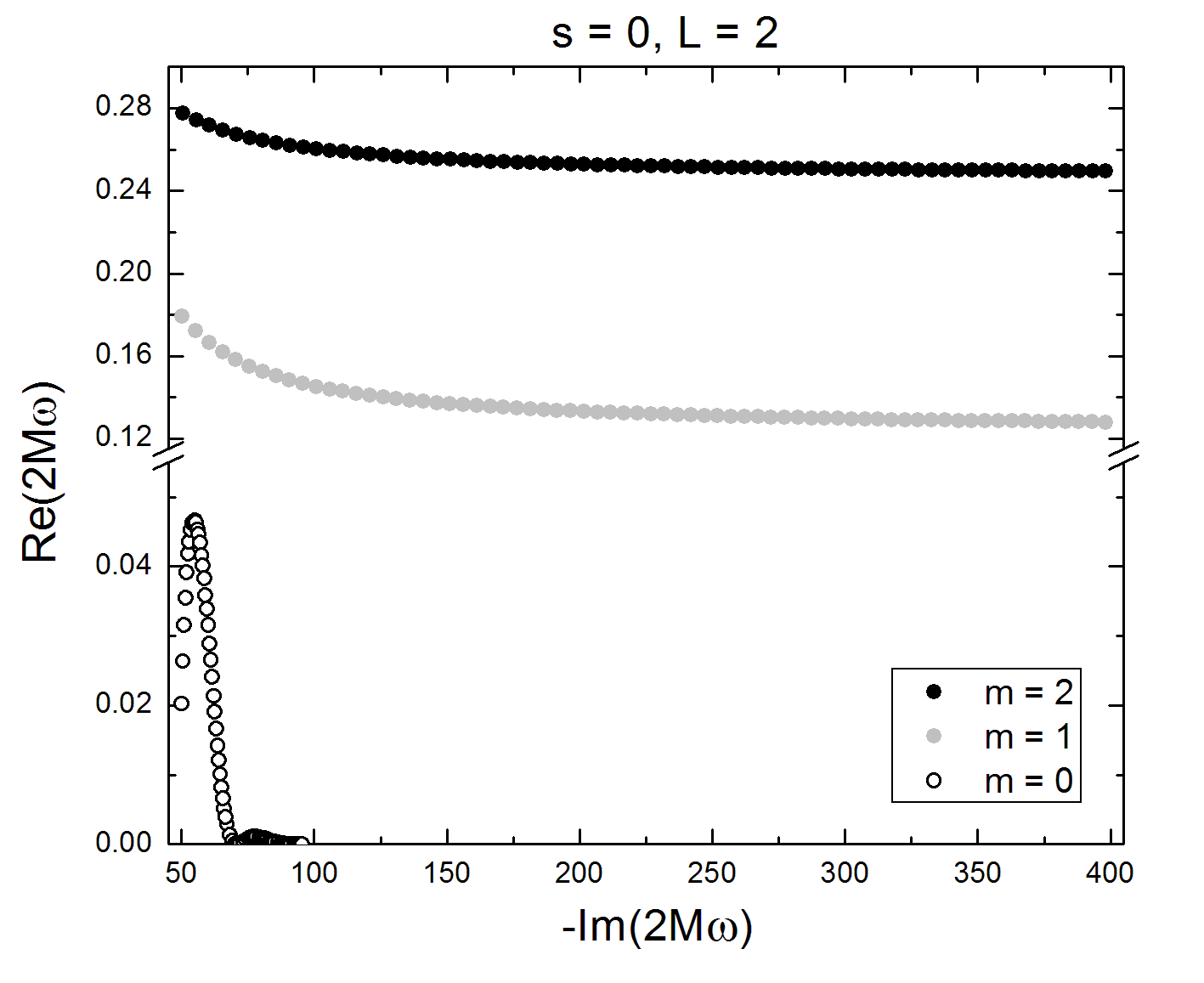}\includegraphics[width=0.45\columnwidth]{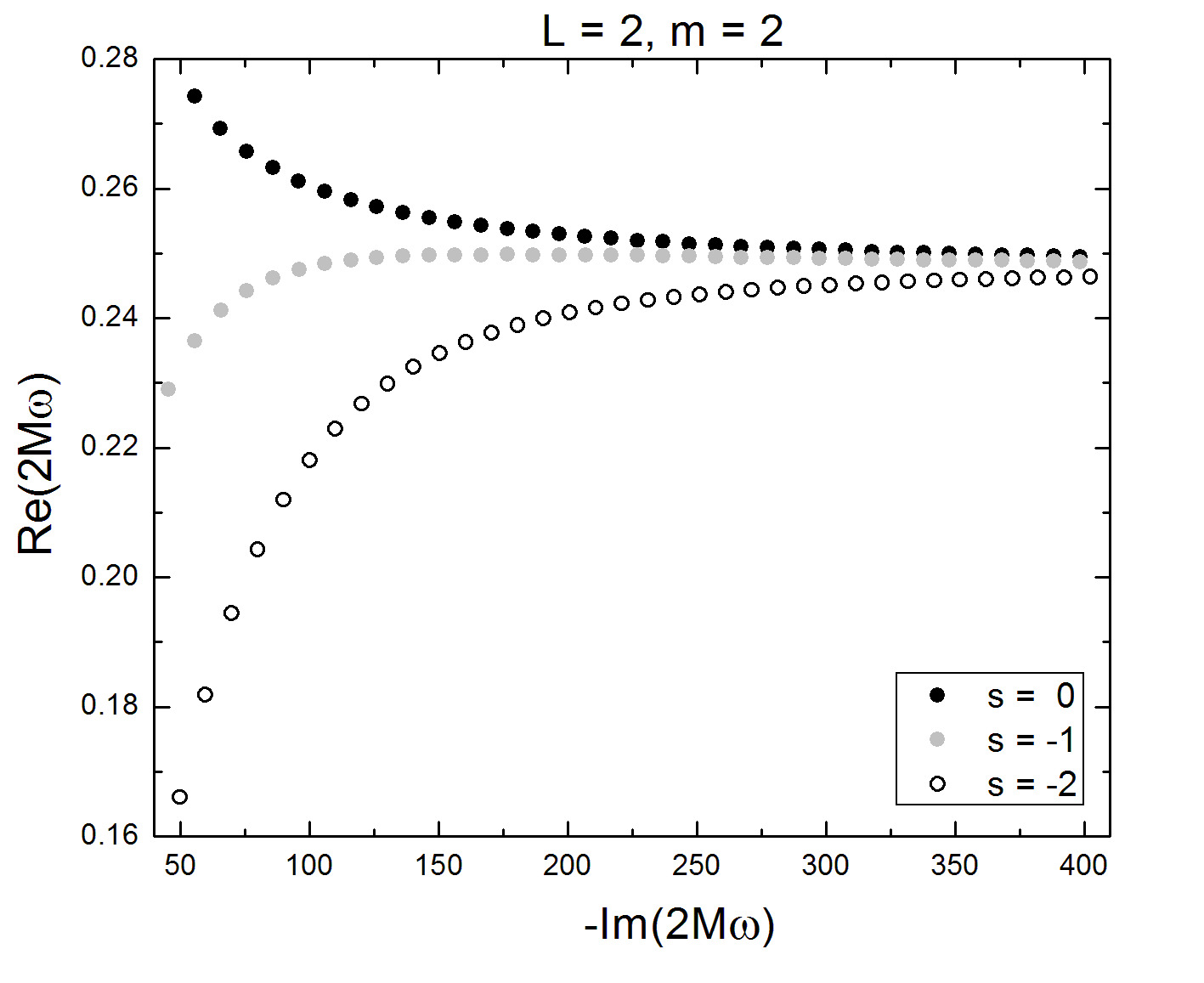}
\par\end{centering}

\caption{Quasinormal mode frequencies for $\ell=2$ modes and $a/M=0.2$ with
$s=0$ and $m$ varied (left) and $m=2$, $s$ varied (right).}
\end{figure}

\begin{figure}
\noindent \centering{}\includegraphics[width=0.45\textwidth]{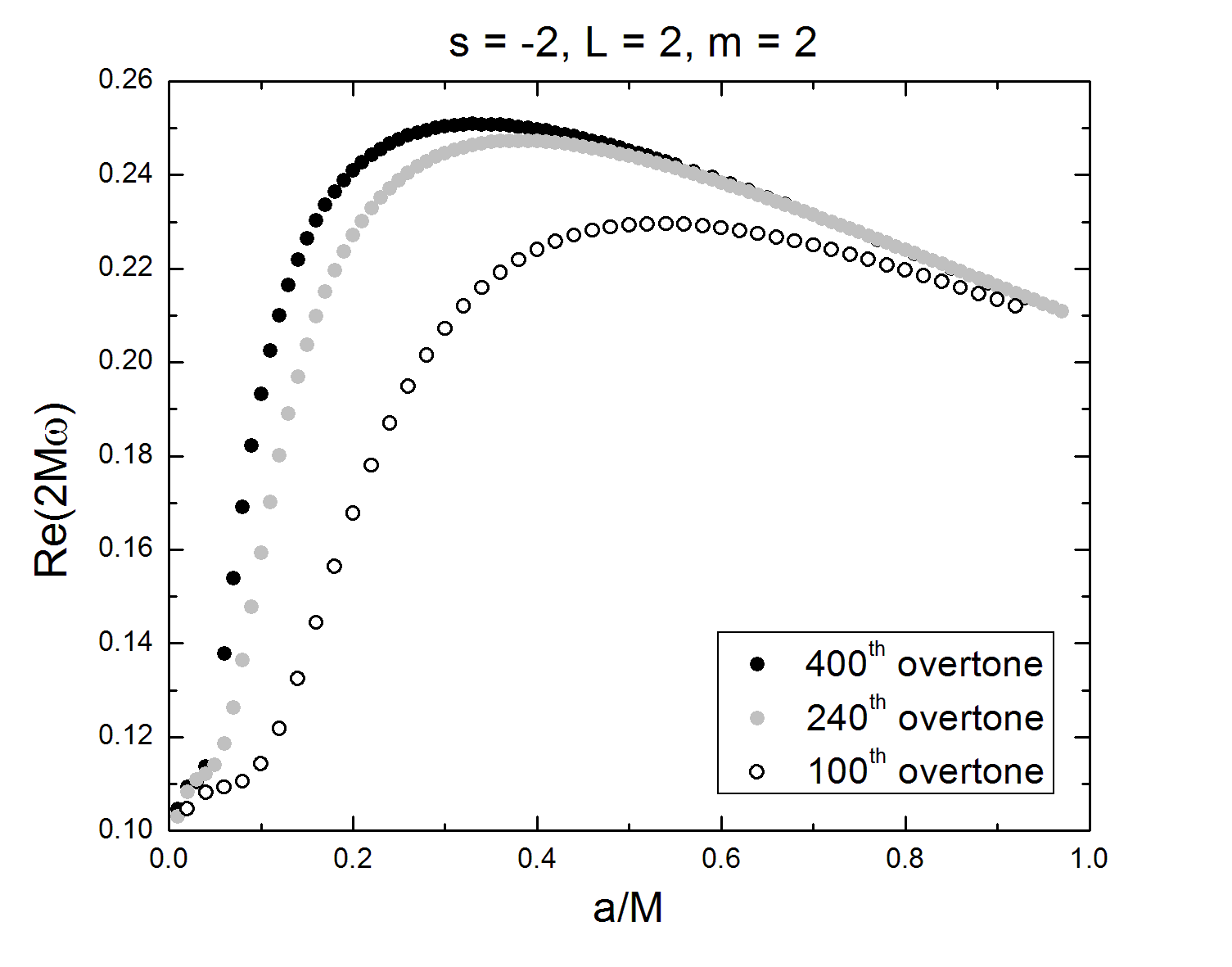}\includegraphics[width=0.45\columnwidth]{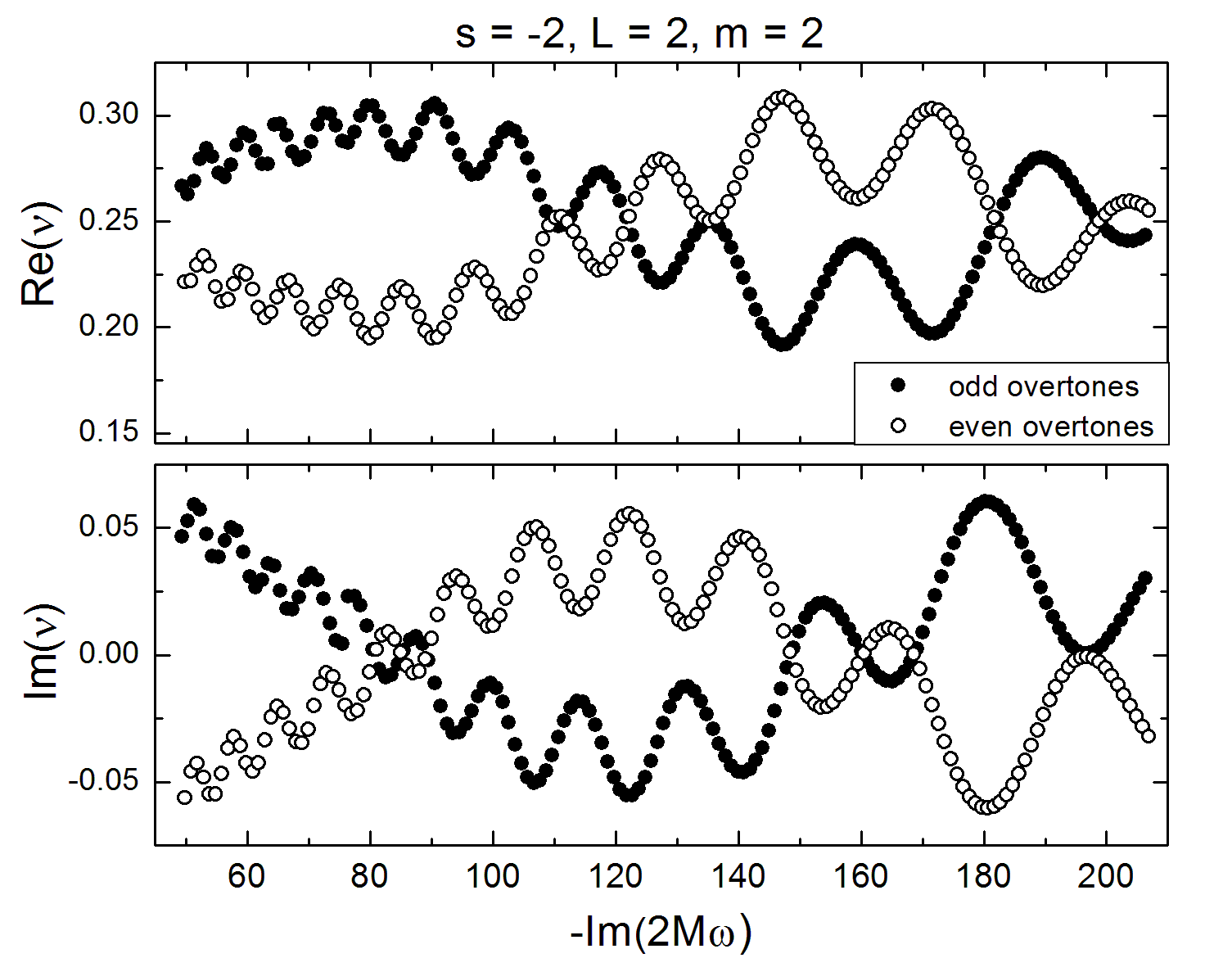}\caption{Numerical values for $s=-2,\,\,\ell=2,\,\, m=2$ modes: real part
of quasinormal mode frequencies for 100th, 240th and 400th overtone
as a function of $a/M$ (left) and real and imaginary $\nu$ values
at high overtones for $a/M=0.2$ (right).}
\end{figure}

\begin{figure}
\noindent \begin{centering}
\includegraphics[width=0.45\textwidth]{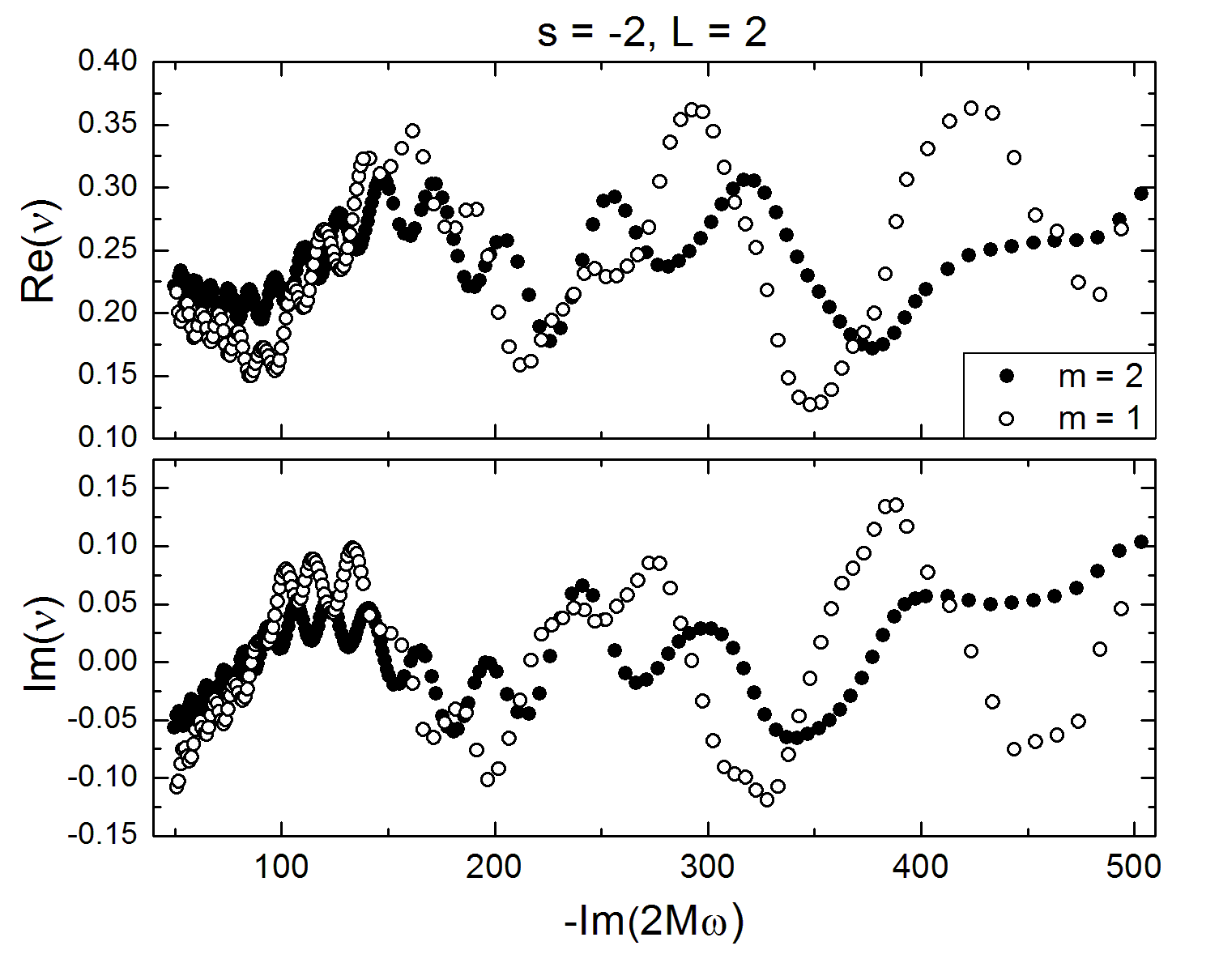}\includegraphics[width=0.45\textwidth]{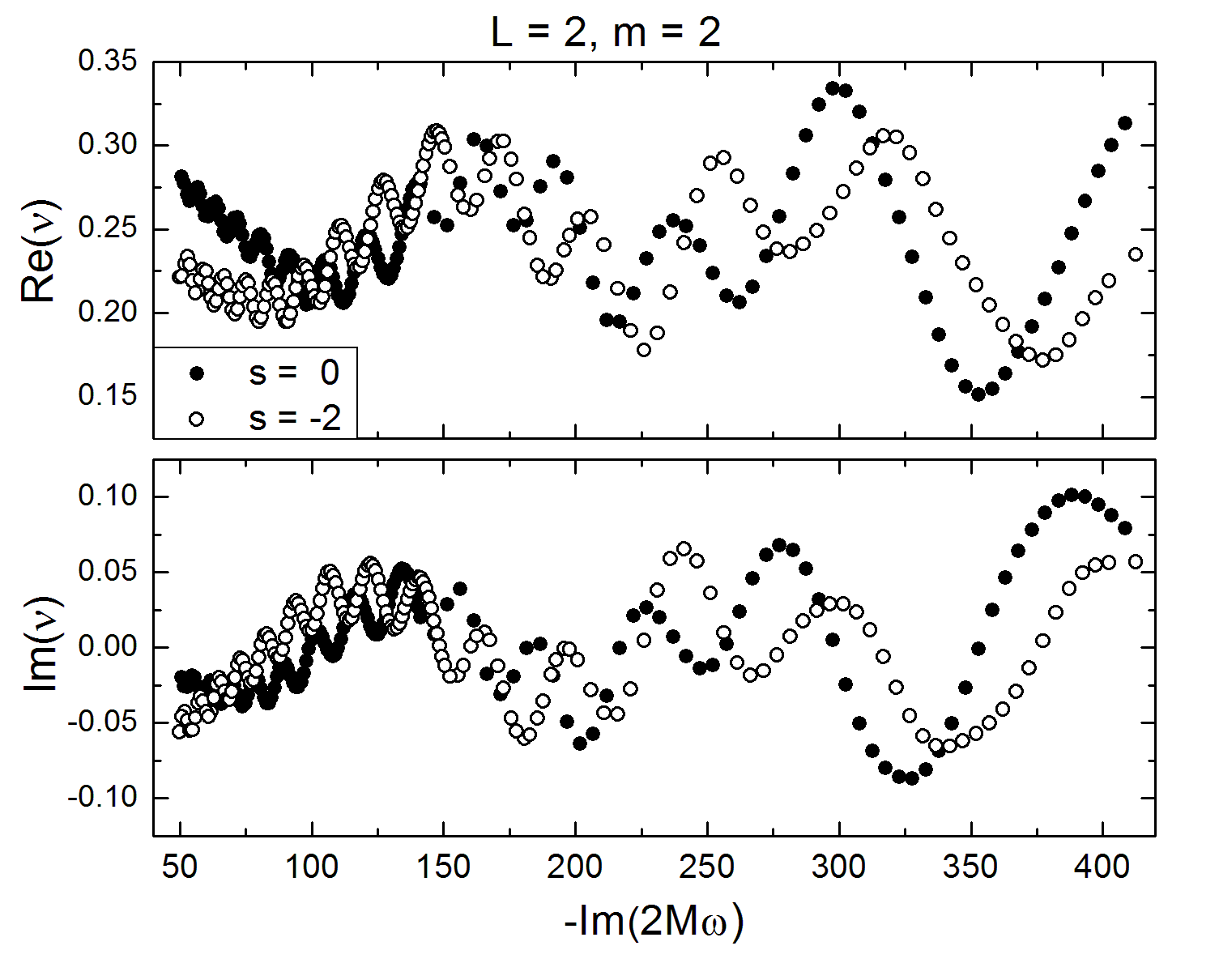}
\par\end{centering}

\caption{Comparison of real and imaginary $\nu$ values at high overtones for
$m=1$ and $m=2$ and fixed $s=2,\,\,\ell=2$ at $a/M=0.2$ (left)
and for $s=0$, $s=-2$ with fixed $\ell=2,\,\, m=2$ (right).}

\end{figure}

\section{Conclusions}

For a general massless excitation in a Kerr black hole background
it is possible to compute a universal function $\nu(\omega,l,m,s)$
which determines a sequence of irreducible representations of the
group $SL(2,\mathbb{R})\times SL(2,\mathbb{R})$ which gives an exact
expression for the mode function. These representations for general
black hole rotation parameter $a$ do not correspond to highest/lowest
weight representations as originally conjectured in the Kerr/CFT literature,
but we have pointed out this feature also arises for the BTZ black
hole. By analogy we may view this as evidence for an underlying CFT
formulation where these modes are related to unitary primary operators.

From the dual holographic field theory, correlators in the bulk are
reproduced by placing the dual CFT at finite left/right moving temperatures.
However to resolve detailed scattering amplitudes from past null infinity
to future null infinity in the Kerr background, it is also necessary
to understand the renormalization group flow from a purported holographic
dual to asymptotically flat spacetime to the Kerr/CFT structure that
takes over at finite values of the radius. 

\appendix

\section*{Appendix - Numerical evaluation of quasinormal frequencies \label{sec:Appendix---Numerical}}

Here we present a brief numerical recipe. The solution to the angular
or radial eigenvalue equation can be expressed as an expansion in
polynomials, with coefficients satisfying a recurrence equation \cite{Leaver:1985ax}
of the form:

\begin{eqnarray*}
\alpha_{0}^{\theta,r}a_{1}^{\theta,r}+\beta_{0}^{\theta,r}a_{0}^{\theta,r} & = & 0\\
\alpha_{n}^{\theta,r}a_{n-1}^{\theta,r}+\beta_{n}^{\theta,r}a_{n}^{\theta,r}+\gamma_{n}^{\theta,r}a_{n-1}^{\theta,r} & = & 0
\end{eqnarray*}
where superscript $\theta,r$ is to denote one or the other equation.
The angular separation constant $E$ and quasinormal mode frequency
$\omega$ can be determined as roots of the corresponding continued
fraction equations:

\begin{equation}
\beta_{0}^{\theta,r}=\frac{\alpha_{0}^{\theta,r}\gamma_{1}^{\theta,r}}{\beta_{1}^{\theta,r}-}\frac{\alpha_{1}^{\theta,r}\gamma_{2}^{\theta,r}}{\beta_{2}^{\theta,r}-}\ldots\frac{\alpha_{n}^{\theta,r}\gamma_{n+1}^{\theta,r}}{\beta_{n+1}^{\theta,r}-}\ldots\label{eq:contfrRT}
\end{equation}

Numerical evaluation of this problem requires truncation to a finite
number of terms. To improve convergence, we implement the Nollert
algorithm \cite{Nollert:1993zz} to evaluate the remainder of the
continued fraction,

\begin{equation}
R_{N}^{\theta,r}\simeq\frac{\gamma_{N+1}^{\theta,r}}{\beta_{N+1}^{\theta,r}-}\frac{\alpha_{N+1}^{\theta,r}\gamma_{N+2}^{\theta,r}}{\beta_{N+2}^{\theta,r}-}\ldots\label{eq:nollertR}
\end{equation}
for some large $N$. For $N\gg1$ the remainder $R_{N}^{\theta,r}$
will be well approximated with a large $N$ expansion, 

\[
R_{N}^{\theta,r}(\omega,A_{lm})\simeq C_{0}^{\theta,r}+C_{1}^{\theta,r}N^{-1/2}+C_{2}^{\theta,r}N^{-1}+\ldots+C_{2k}^{\theta,r}N^{-k}+\ldots\,.
\]

By rewriting \eqref{eq:nollertR} in an implicit form, $R_{N}^{\theta,r}-\gamma_{N+1}^{\theta,r}/(\beta_{N+1}^{\theta,r}-\alpha_{N+1}^{\theta,r}R_{N+1}^{\theta,r})$
and expanding for large $N$, we read off the coefficients $C_{k}^{\theta,r}(\omega,E)$.
For the radial continued fraction we impose $Re(C_{1})>0$ following
\cite{Nollert:1993zz}; for the angular continued fraction all half-integer
powers vanish.

At low overtones, we assume an initial value for $\omega_{0},\, E_{0}$
around which we look for solutions, and evaluate the remainder $R_{N}(\omega_{0},E_{0})$
for a set number of terms where we apply cutoff, $N=n_{\text{cutoff}}$.
We are then in a position to solve \eqref{eq:contfrRT}. We utilize
an iterative procedure, separately computing radial and angular continued
fractions \eqref{eq:contfrRT}, and feeding solutions back into the
next iteration. We increase $n_{\text{cutoff}}$ with each iteration,
until results converge to the desired precision.

At high overtones a much more reliable method is to replace \eqref{eq:contfrRT}
with its inversion \cite{Leaver:1985ax},

\[
\beta_{0}^{r}-\frac{\alpha_{n-1}^{r}\gamma_{n}^{r}}{\beta_{n-1}^{r}-}\frac{\alpha_{n-2}^{r}\gamma_{n-1}^{r}}{\beta_{n-2}^{r}-}\ldots\frac{\alpha_{0}^{r}\gamma_{1}^{r}}{\beta_{0}^{r}}=\frac{\alpha_{n}^{r}\gamma_{n+1}^{r}}{\beta_{n+1}^{r}-}\frac{\alpha_{n+1}^{r}\gamma_{n+2}^{r}}{\beta_{n+2}^{r}-}\ldots
\]
for any positive integer $n$. We find good convergence for the number
of terms in the inverted continued fraction of the order of the overtone
number.
\begin{acknowledgments}
We thank E. Berti for helpful discussions. This work is supported
in part by DOE DE-FG02-13ER42023-Task A and by an FQXi grant.
\end{acknowledgments}
\bibliographystyle{apsrev}
\bibliography{hidden}

\end{document}